# Investigating Transient Characteristics of Volatile Hysteresis and Self-Heating of PrMnO$_3$ based RRAM

J. Sakhuja, S. Lashkare, K. Jana, U. Ganguly

*Abstract*— **PrMnO$_3$ (PMO) based RRAM shows selector-less behavior due to high-non-linearity. Recently, the non-linearity, along with volatile hysteresis, is demonstrated and utilized as a compact oscillator to enable highly scaled oscillatory neurons, which enable oscillatory neuromorphic systems found in the human cortex. Hence, it is vital to understand the physical mechanisms behind such a volatile hysteretic behavior to provide useful insights in developing a device for various neuromorphic applications. In this paper, we present a comprehensive investigation of the transient characteristics and propose a physical mechanism to replicate the observations by simulations. First, we investigate the complex dynamics of the hysteresis with the voltage ramp rate. We observe that the voltage window initially increases and later decreases as the ramp rate is increased - while the current window reduces monotonically. Second, an analytical electrothermal model based on space charge limited current (SCLC) and Fourier Heat equation is proposed to model the co-dependent heat and current flow. Finally, we show that the interplay between the self-heating due to the current and the current dependence on the temperature is accurately modeled to reproduce the hysteresis dependence on the various voltage ramp rate. Such a detailed understanding of the device PMO RRAM volatile hysteresis may enable efficient device design required in neuromorphic computing applications.**

*Index Terms*— **PMO, RRAM, Self-heating, Thermal Feedback, Volatile Hysteresis**

## I. INTRODUCTION

Brain-inspired neural architectures have been the subject of extensive research for alternate computing paradigms. Previous studies have shown that the human brain seamlessly processes data parallelly and in an energy-efficient manner [1]. In order to incorporate these strengths of brain functioning with our technologies, Oscillatory Neural Networks (ONNs) are gaining interest with researchers[2]–[4]. Networks of coupled oscillators have been shown to exhibit associative memory and solve optimization problems[5]. Although there are many CMOS based oscillator circuits[2] available, which can be utilized in ONNs, relaxation oscillators based on single scalable devices are more desirable for area and energy efficiency in large network sizes. The key element of an oscillator is a hysteretic DC IV characteristic, as in a Schmitt trigger. Many device level oscillators in different material systems and device physics like VO2 [6] and NbO2 (insulator to metal transition) [7], [8], TaOx (conductive volatile filament)[9], and spintronics based devices have been proposed[10].

Recently PrMnO3 (PMO) based RRAM device has demonstrated highly non-linear I-V characteristics with threshold switching and a volatile hysteresis loop in Low Resistance State (LRS), enabling a compact oscillator[11], [12]. An efficient implementation of the Rectified Linear Unit (ReLU) type neuron, which mimics the integration function in neuron using the sharp current shoot up in the current transient in the PMO device, has been shown[13]. These devices are highly scalable, non-filamentary, and demonstrate excellent endurance and retention [14], [15]. Further, the existence of oscillatory (volatile switching) and memory (non-volatile RRAM and neural synapse) functionalities in the same material systems render these devices as the device of choice for large scale integrated ONNs and Neuromorphic computing. Though extensive applications of the hysteresis have been demonstrated experimentally, the physical phenomena governing the dynamics of the critical hysteresis window formation are yet to be explored. This is critical as the PMO device in an oscillator circuit swings between its volatile switching hysteretic thresholds. Depending on oscillation frequency, the effective rate of voltage sweep (ramp rate) across the device can be different. Hence, the ramp rate dependence of the hysteresis window is crucial to accurately predict the nature of oscillations at different frequencies.

In this work, we present the hysteresis experimentally by varying the voltage ramp rate to observe complex behavior. To understand the observations quantitatively, an analytical electro-thermal model is proposed to simulate the current-temperature transient. The model effectively captures the interplay between current, temperature, and voltage ramp rate.

## II. DEVICE FABRICATION

The stack of PMO based RRAM is as shown in Fig. 1. The Silicon (Si) wafer with <100> orientation is used for RRAM fabrication. A SiO$_2$ layer of 700 nm thickness is thermally grown on Si wafer by wet oxidation process using 4" Wet Oxidation Furnace. A bilayer of Titanium (Ti) followed by Platinum (Pt) is deposited on SiO2 through DC sputtering. The Ti layer provides adhesion between Pt and SiO2. The Pt forms the bottom metal contact layer of the RRAM structure.

The work is partially funded by Department of Science and Technology (DST), Nano Mission and Ministry of Electronics and IT (MeitY), India. It was performed at IIT Bombay Nanofabrication Facility. All authors are with Electrical Engineering, IIT Bombay, Mumbai, India-400076. S. Lashkare is supported jointly by the Visvesvaraya Ph.D. Scheme of MeitY, Government of India, being implemented by Digital India Corporation and Intel Ph.D. Fellowship. E-mail: udayan@ee.iitb.ac.in

Following this, a blanket layer of 60 nm PMO film is deposited using RF sputtering at room temperature. The PMO film is amorphous in an as-deposited state. In order to crystallize the PMO film, annealing is done in the controlled environment of $N_2 : O_2$ partial pressures (95:05) at 750°C for 30 seconds[16]. Further, through UV photolithography, different top contact areas are patterned on PMO film. Finally, tungsten(W) top metal contact is formed through the DC sputtering and metal lift-off process. The square area devices with 10μm × 10μm dimensions are used for the study.

For all the DC I-V measurements of the fabricated devices, Agilent B1500A Semiconductor Analyzer is used, whereas for transient measurements B1530 waveform generator/fast measurement unit (WGFMU) is used. All the measurements are done at room temperature.

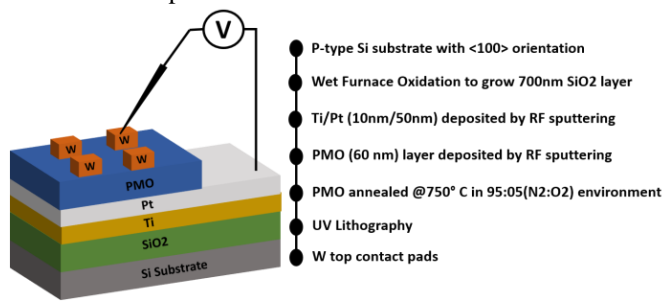

Fig. 1 Device Schematic of fabricated PMO based RRAM and Fabrication process flow

III. RESULTS AND DISCUSSION

*A. DC IV characteristics*

PMO is a p-type semiconductor material, sandwiched between two metal electrodes (Pt - bottom and W - top). The voltage (V) is given as the input to the top electrode while the bottom electrode is grounded, and current (I) is measured as the output. When a DC voltage sweep from 0 to -1.5 V is applied to the device, at a certain high threshold voltage ($V_H$) the device switches its current resistance state ($R_H$) abruptly to a low resistance state ($R_L$) (Fig. 2). The full voltage sweep is as shown in the inset of Fig. 2. On the application of voltage sweep from -1.5 V to 0 V, the device switches back from $R_L$ to $R_H$ at a lower threshold voltage ($V_L$) forming a hysteresis memory window. The abrupt rise in current is attributed to self-heating[15], [17]–[20] within the PMO material due to its low thermally conductive nature (300 × lower as compared to Silicon (1.48Wcm$^{-1}$K$^{-1}$) [21]. When the voltage is applied, current starts flowing through the device resulting in an increase in the device temperature through Joule heating. The low thermally conductive nature of PMO results in an accumulation of heat that causes an increase in temperature within the device which further aids the increase in current. At a certain threshold voltage, the positive feedback between current and temperature results in a sharp shoot-up in the current [11],[13]. Here, the voltage sweep does not incorporate much information as the current switching is abrupt. Hence, a current sweep (current is injected in the device, and the voltage is measured) is performed to capture the gradual changes in the device. The decrease in resistance due to high device temperatures i.e. Negative Differential Resistance (NDR) can be seen in current sweep characteristics (Fig. 2). The NDR region is an S-shaped profile of the IV characteristics indicating a decrease in voltage with increase in current. The current compliance is set so as to avoid device breakdown. The hysteresis window is limited by current compliance.

As stated earlier, a strong correlation exists between current conduction and heat dissipation within the devices. Since temperature changes have an associated timescale, the nature of hysteresis is expected to change with the voltage (V) ramp rate. While DC behavior is well-known, oscillations utilize the hysteresis in a transient mode [12]. Hence the transient response to various V ramp rates is essential to investigate - which is presented next.

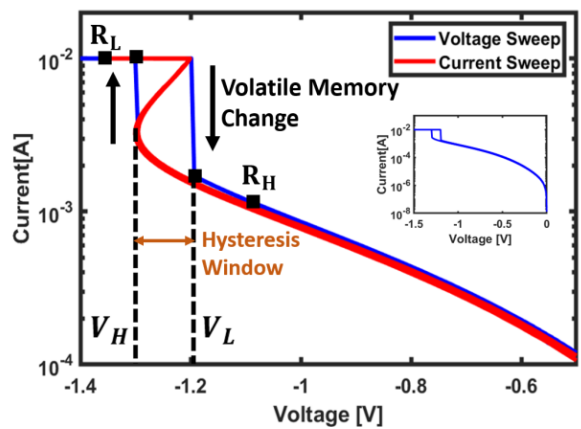

Fig. 2 Hysteretic DC I-V of PMO device showing a sharp rise and fall in current after a voltage threshold $V_H$ and $V_L$ respectively (blue curve). This represents the volatile change between two resistance states $R_H$ and $R_L$. A hysteresis window is formed between two threshold voltages. Voltage sweep showing the sharp shoot-up (High NL) at high threshold voltage ($V_H$). The current sweep shows the S-shaped characteristics starting at the $V_H$ threshold voltage that indicates a current-controlled negative differential region (NDR) region.

*B. Ramp rate-dependence of $I(V)$ characteristics*

The voltage ramp rate is described as the rate of change in the voltage (ΔV) with respect to change in time (Δt). For the following experiments, ΔV is kept constant, and Δt is varied to obtain different ramp rates. A transient voltage sweep of varying ramp rates was applied, and current through the device is measured. The step voltage (ΔV) of 20 mV was used for different voltage ramp rates, as shown in the inset of Fig. 3. The voltage of -1.4V is taken as maximum voltage to be applied during transient measurements as it exceeds high threshold voltage ($V_H$) in DC IV characteristics (Fig. 2).

In Fig. 3, we observe that there is a complex dependence of varying ramp rates on the width of the hysteresis window. Specifically, we observe a non-monotonic trend indicated by an initial increase in width of the hysteresis window with a decrease in ramp rate, whereas for sufficiently low ramp rates, the width decreases owing to fixed maximum voltage. Additionally, a monotonic increase in the current peak and a monotonic decrease in switching threshold voltage with a decrease in ramp rate can also be observed.

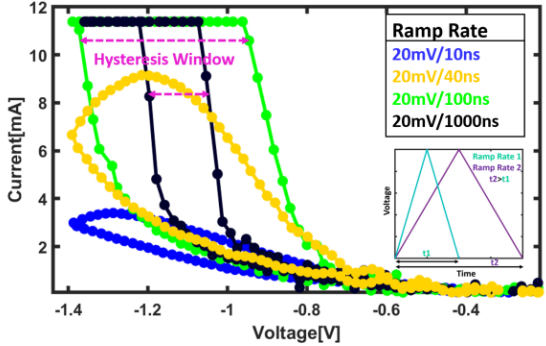

Fig. 3 Experimental transient characteristics of PMO based RRAM devices, demonstrating current evolution with voltage for different voltage ramp rates. Very fast ramp rate (blue) leads to no hysteresis. Further Increasing ramp rate, hysteresis starts to develop (yellow), then increases (green) and finally decreases (black). The inset shows methodology for different Ramp rates. Given a fixed voltage step, the ramp rate varies with varying time steps. The 'Ramp Rate 2' is smaller than 'Ramp rate 1'. The ramp rate represents the slope of voltage versus time plot.

Next, the current transient response to ramp voltage for varying ramp rates is shown in Fig. 4. For a fast ramp rate ($\Delta t = 10 ns$), the current resembles the voltage ramp, i.e., approximately triangular (Fig. 4a). The peak current also occurs close to the peak voltage. Peak currents are also low (~$4 mA$). As the ramp rate is reduced ($\Delta t = 40 ns$), the current becomes increasingly non-linear (Fig. 4b). The peak current increases to ~$10 mA$ and occurs after the peak voltage. For slower ramp ($\Delta t = 100 ns$), the non-linear increase in current becomes sharper (Fig. 4c). The current increases and reaches compliance at the time close to the peak of the voltage ramp. For an even slower ramp ($\Delta t = 1000 ns$), the non-linear increase in current becomes even sharper (Fig. 4d). The current increases and reaches compliance relatively much earlier than the peak of the voltage ramp.

In Fig. 5, the timescales for triangular voltage ramps are shown in log scale. We can clearly note that the response peak current increases with increase ramp time ($\Delta t$) and saturates due to compliance.

*C. Discussion*

Given the complex ramp dependence of hysteresis observed, a qualitative explanation is of significant interest. For a slow voltage ramp rate, i.e., with sufficiently long $\Delta t$ time step, current at that voltage will tend to its steady-state value. If there is self-heating within the device, the temperature will also attain a steady-state value. However, for shorter $\Delta t$, the voltage is changed before the current or temperature could attain its steady-state value. The effect will propagate with every time step. Thus, current and temperature will lag the steady-state value as ramp rate increases, i.e., $\Delta t$ decreases. For a fast ramp rate, i.e., $\Delta t$ is small (10 ns), the current curve traces the instantaneous voltage with the peak of current and voltage occurring coincidentally in time. The current levels are adequately low at that point (Fig.4a). As it has been discussed earlier, in PMO based devices, current-temperature feedback is an additional contributor to the current. Low current levels with a fast ramp rate indicate that the current-temperature positive feedback has not yet initiated. As ramp rate decreases, i.e., $\Delta t$ increases, the current becomes increasingly non-linear, which indicates the enhanced effect of current and temperature positive feedback getting more time to enhance non-linearity. Thus, the current increase with reduced ramp rates, consistent with Fig 5. In comparison to the monotonic current dependence on ramp rate, the non-monotonic voltage(V) hysteresis is more interesting – which we discuss next. At steady state, the $V_H$ and $V_L$ maybe some value. However, as the ramp rate increases, the non-linear current transition "lags" the voltage for both ramp up and ramp down. So, the $V_H$ increases and $V_L$ decreases. Thus, initially, V-hysteresis should increase – consistent with our observation. However, at every faster ramp rate, the peak current reduces (Fig. 5), and the current hysteresis also reduces (Fig. 3). Ultimately, for the fastest ramp rate ($\Delta t = 40 ns, 10 ns$), minimum current hysteresis is observed. Thus, the V-hysteresis will eventually reduce – producing the complex non-monotonic dependence – where it initially increases and then decreases with ramp rate – consistent with Fig. 3.

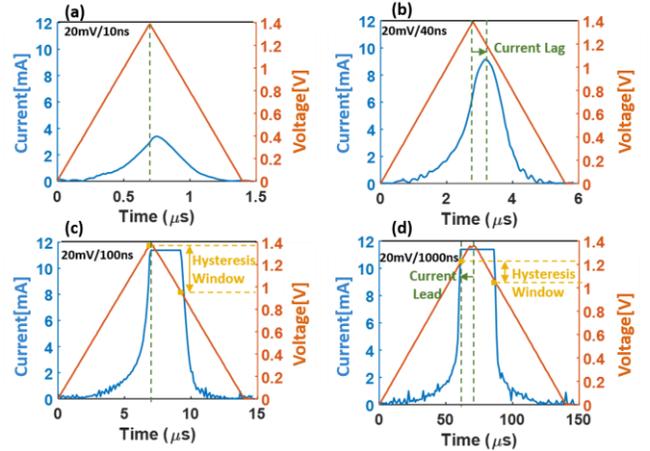

Fig. 4 Experimental transient characteristics of PMO based RRAM devices, demonstrating current and voltage evolution with time for different voltage ramp rates. The ramp rate of (a) 20 mV/10ns, current peak occurring coincidentally in time with voltage peak (b) 20 mV/40 ns, current peak lagging voltage peak, (c) 20 mV/100ns current peak occurring coincidentally in time with voltage peak and current hitting the compliance, (d) 20 mV/1000ns current peak leading voltage peak.

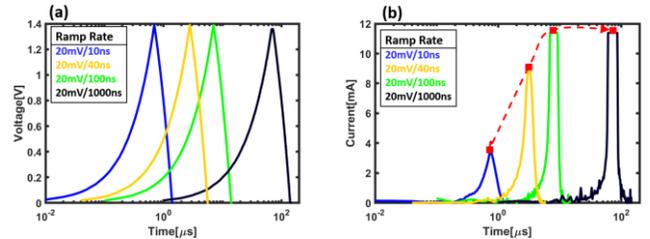

Fig. 5 Experimental (a) Applied voltage bias with different ramp rates, (b) Current measured corresponding to the voltage applied with different ramp rates.

*D. Electrothermal Model for transient IV hysteresis*

In this section, we demonstrate an electrothermal model to quantitatively show the V-ramp dependence of the hysteresis to further substantiate our qualitative model. We have earlier demonstrated Space Charge Limited Current (SCLC) conduction mechanism in PMO based RRAM devices [22]. Further, the highly non-linear switching characteristics demonstrated by PMO material are associated with self-heating and thermal feedback mechanism within the device [17]. Here we use an SCLC current conduction based thermal feedback model (Fig. 6) for transient simulations to validate the experimental results. To model temperature dependence on time, the voltage is input as a ramp function with varying ramp rates. The current and device temperature is obtained as outputs from the simulations. The model used SCLC to compute charge current where Temperature and Voltage is an input. As the resistance change is volatile, there is no need to consider any ionic motion or trap density change. Further, the temperature is calculated by Fourier Heat Equation (4).

$$I = I_{Ohmic} + I_{SCLC} \qquad (1)$$

$$I_{Ohmic} = qA\mu N_v \left(\frac{T}{T_{amb}}\right)^{3/2} e^{\left(-\frac{q\phi_B}{kT}\right)} \left(\frac{V}{L}\right) \qquad (2)$$

$$I_{SCLC} = A\mu\epsilon_o\epsilon_{PMO} \left(\frac{N_v}{N_T}\right) \left(\frac{T}{T_{amb}}\right)^{3/2} e^{\left(-\frac{qE_{trap}}{kT}\right)} \left(\frac{V^2}{L^3}\right) \qquad (3)$$

where $\mu$(Mobility)=3cm$^2$/Vs [20], $\phi$(Barrier Height)=0.05 eV, $\epsilon_{PMO}$(Permittivity)=30 [23], $N_v$ is Effective density of states, $E_{trap}$ is Trap energy level, $N_T$ is Trap density, L is Thickness of PMO = 60 nm, $T$ is Temperature of Device, $T_{amb}$ (Ambient Temperature)=300K

$$-k\frac{d^2T}{dx^2} + c_v\frac{dT}{dt} = \frac{IV}{\text{volume}} \qquad (4)$$

where k is Thermal conductivity, $c_v$ is Specific heat capacity, volume = A. L = $10 \times 10\mu m^2 \times 60nm$

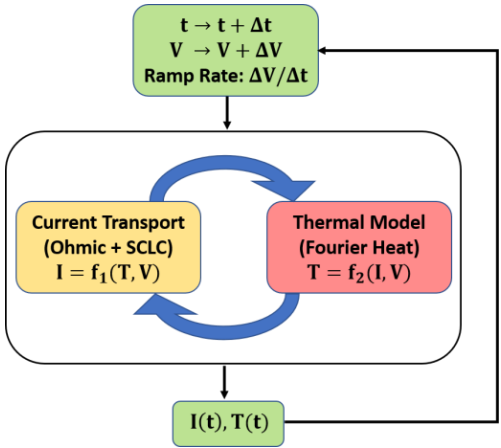

Fig. 6 Thermal Feedback model incorporating SCLC current Transport and Fourier Heat transport to capture self-heating based transient dynamics of current temperature with voltage.

The simulated results for current transient with voltage for different ramp rates are shown in Fig 7. As seen in Fig. 3 and 7, the experimental and simulation IV characteristics are in good agreement. The model produces a minimum current hysteresis at high ramp rates. At lower ramp rates, the current hysteresis increases, which leads to a volatile hysteretic memory window. Thus, our model that voltage ramp rates regulate the positive thermal feedback within the device quantitatively reproduces experimental observations.

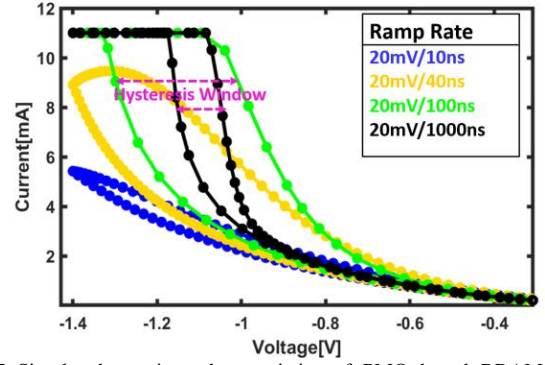

Fig 7 Simulated transient characteristics of PMO based RRAM devices, demonstrating current and voltage evolution with time for different voltage ramp rates.

Additionally, the effective temperature within the device with respect to ambient temperature is extracted from device simulation. Fig. 8 shows temperature evolution with time, and it is compared with the current. For fast ramp rates, when the current levels are low, the temperature within the device is also low. Temperature follows the current trends due to Joule heating, i.e., temperature increases with an increase in current and vice a verse but with a lag. A shift in temperature curve and current curve can be observed. It indicates a relative lag in temperature with respect to current increases for faster ramp rates, and the maximum temperatures achieved are lower. With a decrease in ramp rate, a rise in temperature becomes more prominent, and the lag in both the current and temperature transient decreases. Essentially, due to the slow ramp rate, the temperature gets enough time to rise towards the steady-state value and hence facilitates increased levels of current. Thus, the agreement of our model to experiment strongly supports that the voltage ramp rates regulate the positive thermal feedback within the device quantitatively reproduces experimental observations. Hence, thermal properties (i.e., heating/cooling timescales) engineering will enable the control of temperature transient during V-ramp to enable the design of hysteresis, which is a critical aspect of oscillator design.

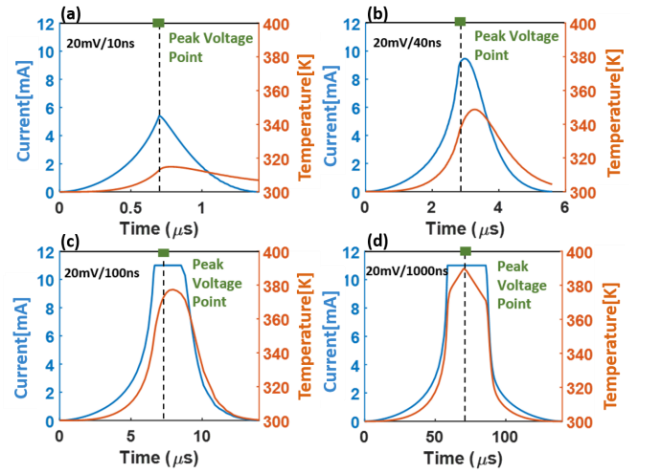

Fig 8 Simulated transient characteristics of PMO based RRAM devices, demonstrating current and temperature evolution with time for different voltage ramp rates, (a) The ramp rate of 20 mV/10ns. The temperature build-up within the device is low, (b) The ramp rate of 20 mV/40ns. Temperature increases with time, (c) Ramp rate of 20 mV/100ns. Temperature increases further, and current hits the compliance, (d) The ramp rate of 20 mV/1000ns. Temperature closely follows the current, demonstrating a strong mutual dependence.

## IV. Conclusion

In summary, this work establishes a strong dependence of hysteretic device characteristics on the device voltage ramp rate. The voltage ramp rate controls the device temperature, which indeed determines the device's volatile resistive switching speeds. The temperature and current transients result in hysteresis memory window due to the thermal feedback mechanism. The proposed thermal feedback model-based simulation results corroborate the experimental findings. Thus, self-heating-based switching characteristics are important from the design prospect of high-speed oscillators to be utilized in large scales ONNs.